\documentclass[12pt]{amsart}

\usepackage{amsmath,amssymb,epic,lscape}

\newtheorem{theorem}{Theorem}[section]
\newtheorem{proposition}[theorem]{Proposition}

\theoremstyle{definition}

\theoremstyle{remark}

\numberwithin{equation}{section}
\providecommand{\bysame}{\leavevmode\hbox to3em{\hrulefill}\thinspace}

\begin{document}

\def\DJ{{\hbox{D\kern-.8em\raise.15ex\hbox{--}\kern.35em}}}
\def\DJo{$\;$\kern-.4em
    \hbox{D\kern-.8em\raise.15ex\hbox{--}\kern.35em okovi\'c}}

\def\NSERC{Supported in part by an NSERC Discovery Grant.}

\font\germ=eufm10
\def\sL{{\mbox{\germ sl}}}

\def\al{{\alpha}}
\def\be{{\beta}}
\def\bt{{\bullet}}
\def\kr{{\circ}}
\def\vf{{\varphi}}
\def\la{{\lambda}}
\def\bR{{\mbox{\bf R}}}
\def\bZ{{\mbox{\bf Z}}}
\def\bC{{\mbox{\bf C}}}
\def\bN{{\mbox{\bf N}}}
\def\bH{{\mbox{\bf H}}}
\def\pA{{\mathcal A}}
\def\pF{{\mathcal F}}
\def\pE{{\mathcal E}}
\def\pC{{\mathcal C}}
\def\pX{{\mathcal X}}
\def\pY{{\mathcal Y}}
\def\pZ{{\mathcal Z}}
\def\Cb{{\bar{C}}}
\def\tr{{\rm tr\;}}
\def\Tr{{\rm Tr\;}}
\def\impl{\ \Rightarrow\ }
\def\Aut{{\mbox{\rm Aut}}}
\def\Sk{{\mbox{\rm Skew}}}
\def\GL{{\mbox{\rm GL}}}
\def\SL{{\mbox{\rm SL}}}
\def\SO{{\mbox{\rm SO}}}
\def\Sp{{\mbox{\rm Sp}}}
\def\Un{{\mbox{\rm U}}}
\def\Ort{{\mbox{\rm O}}}

\renewcommand{\subjclassname}{\textup{2000} Mathematics Subject
Classification }

\title[Periodic complementary binary sequences]
{ Periodic complementary sets of binary sequences }

\author[D.\v{Z}. \DJ okovi\'{c}]
{Dragomir \v{Z}. \DJ okovi\'{c}}

\address{Department of Pure Mathematics, University of Waterloo,
Waterloo, Ontario, N2L 3G1, Canada}

\email{djokovic@uwaterloo.ca}

\thanks{\NSERC}


\date{}

\begin{abstract}
Let $PCS_p^N$ denote a set of $p$ binary sequences of 
length $N$ such that the sum of their periodic
auto-correlation functions is a $\delta$-function.
In the 1990, B\"{o}mer and Antweiler addressed the problem of
constructing $PCS_p^N$. They presented a table covering the
range $p\le12$, $N\le50$ and showing in which cases it was
known at that time whether $PCS_p^N$ exist, do not exist,
or the question of existence is undecided. The number of
undecided cases was rather large.

Subsequently the number of undecided cases was reduced to 26
by the author. In the present note, several cyclic difference
families are constructed and used to obtain new sets of periodic
binary sequences. Thereby the original problem of B\"{o}mer and 
Antweiler is completely solved.
\end{abstract}

\maketitle
\subjclassname{ 05B20, 05B30 }
\vskip5mm

\section{Introduction}

Let $a=a(0),a(1),\ldots,a(N-1)$ be a binary sequence of length $N$.
By this we mean that each $a(i)=\pm1$.
The periodic and non-periodic auto-correlation functions
(PACF and NACF) of $a$ are defined by
\[ \tilde{\vf}_a(i)=\sum_{j=0}^{N-1} a(j)a(i+j \text{ mod } N),
\quad 0\le i<N, \]
and
\[ \vf_a(i)=\sum_{j=0}^{N-1-i} a(j)a(i+j),
\quad 0\le i<N, \]
respectively. By convention, $\vf_a(i)=0$ for $i\ge N$, and
$\vf_a(-i)=\vf_a(i)$ for all $i$'s.

A family of $p$ binary sequences $\{a_i\}$, $1\le i\le p$, all
of length $N$, is a family of periodic resp. aperiodic
complementary binary sequences ($PCS_p^N$ resp. $ACS_p^N$) if
the sum of their PACF resp. NACF is a $\delta$-function.
As $\tilde{\vf}_a(i)=\vf_a(i)+\vf_a(N-i)$ for $0\le i<N$,
we have $ACS_p^N\impl PCS_p^N$.

B\"{o}mer and Antweiler \cite{BA2} addressed the problem of
constructing $PCS_p^N$. They presented a table covering the
range $p\le12$, $N\le50$ and showing in which cases it was
known at that time whether $PCS_p^N$ exist, do not exist,
or the question of existence is undecided. The number of
undecided cases was rather large. In our note \cite{DZ2}
we have reduced the number of undecided cases to 26, see
also \cite{FSX}.

In this note we shall construct $PCS_p^N$ covering all
undecided cases.
The construction uses the approach via difference families,
also known as supplementary difference sets (SDS).
The connection is recalled in section \ref{SDS}.

In our preprint \cite{DZ5} we have introduced a normal form
for SDS's. All SDS's presented in the remainder of this note
are written in that normal form. All of them
were constructed by using our genetic type algorithm.

\section{Base sequences and the case $p=4$}

Base sequences, originally introduced by Turyn \cite{RT},
are quadruples $(a;b;c;d)$ of binary sequences, with
$a$ and $b$ of length $m$ and $c$ and $d$ of length $n$,
and such that the sum of their NACF's is a $\delta$-function.
We denote by $BS(m,n)$ the set of such sequences.
According to \cite[p. 321]{HCD} the $BS(n+1,n)$ exist
(we say ``exist'' instead of ``is non-empty'')
for $0\le n\le35$. In our paper \cite{DZ4}
one can find an extensive list of $BS(n+1,n)$
covering the range $n\le32$.

There is a map
\begin{equation} 
BS(m,n)\to ACS_4^{m+n}
\end{equation}
defined by $(a;b;c;d)\to(a,c;a,-c;b,d;b,-d)$, where
$a,c$ denotes the concatenation of the sequences $a$ and $c$,
and $-c$ denotes the negation of the sequence $c$, i.e.,
we have $(-c)(i)=-c(i)$ for all $i$'s. In particular,
for $m=n=N$ we have a map $ACS_4^N=BS(N,N)\to ACS_4^{2N}$.
It follows that $ACS_4^N$ exist for $N\le72$.
Since $ACS_p^N\impl PCS_p^N$, we have

\begin{proposition} \label{slu-4}
$ACS_p^N$ and $PCS_p^N$ exist if $p$ is divisible by $4$
and $N\le72$.
\end{proposition}

\section{Supplementary difference sets} \label{SDS}

Let $\bZ_N=\{0,1,\ldots,N-1\}$ be the cyclic group of order $N$
with addition modulo $N$ as the group operation. For $m\in\bZ_N$
and a subset $X\subseteq\bZ_N$ let $\nu(X,m)$ be the number
of ordered pairs $(i,j)$ with $i,j\in X$ such that
$i-j\equiv m$ $\pmod{N}$. We say that the subsets
$X_1,\ldots,X_p\subseteq\bZ_N$ are supplementary difference sets
(SDS) with parameters $(N;k_1,\ldots,k_p;\la)$ if $|X_i|=k_i$
for all $i$ and
\[ \sum_{i=1}^p \nu(X_i,m)=\la,\quad \forall m\in\bZ_N\setminus\{0\}. \]
If also $p=1$ then $X_1$ is called a difference set.

If $\{a_i\},\,1\le i\le p$, are $PCS_p^N$, then the sets
\begin{equation} \label{uslov}
X_i=\{j\in\bZ_N: a_i(j)=-1\},\quad 1\le i\le p,
\end{equation}
are SDS with parameters $(N;k_1,\ldots,k_p;\la)$, where
$k_i=|X_i|$ for all $i$'s. Moreover, if $N>1$, the following
condition holds:
\begin{equation} \label{red}
4(k_1+\cdots+k_p-\la)=pN.
\end{equation}
The converse is also true:
If $X_1,\ldots,X_p$ are SDS with parameters $(N;k_1,\ldots,k_p;\la)$
satisfying the above condition, then the binary sequences
$\{a_i\},\,1\le i\le p$, defined by (\ref{uslov}) are  $PCS_p^N$.
These facts are easy to prove, see e.g. \cite{AX,BA2}.

One can also show easily that if $(N;k_1,\ldots,k_p;\lambda)$ are
parameters of an SDS then
\[ pN=\sum_{i=1}^p (N-2k_i)^2. \]
This is useful in selecting the possibilities for the parameter
sets of a hypothetical SDS.

\section{The cases $p=1$ and $p=2$}

It follows from (\ref{red}) that if $PCS_1^N$ exists then $N$ is
divisible by 4. The sequence $+,+,+,-$ is a $PCS_1^4$.
No $PCS_1^N$ are known for $N>4$. In fact it is known
(see \cite{BS}) that they do not exist for $4<N<10^{12}$.

The $ACS_2^N$ are known as Golay pairs of length $N$.
We say that $N$ is a Golay number if $ACS_2^N$ exist.
It is known that if $N>1$ is a Golay number then $N$ is even
and not divisible by any prime congruent to 3 $\pmod{4}$.
In particular, $N$ is a sum of two squares.
The Golay numbers in the range $N\le50$ are
1,2,4,8,10,16,20,26,32 and 40. See \cite{BF,DZ3}
for more details and additional references.

If $PCS_2^N$ exist and $N>1$ then (\ref{red}) implies that
$N$ must be even. It is also well known that $N$ must be a sum of
two squares, see e.g. \cite{AX}.
Apart from the Golay numbers, the integers satisfying these
conditions and belonging to the range $N\le50$ are
18,34,36 and 50. It is known that $PCS_2^{18}$ and $PCS_2^{36}$
do not exist \cite{AX,CY}. Three non-equivalent examples of
$PCS_2^{34}$ and a single example of a $PCS_2^{50}$
have been constructed in our papers \cite{DZ1,DZ2,DZ5}.

In particular the following holds
\begin{proposition} \label{slu-2}
In the range $N\le50$, $PCS_1^N$ exist iff $N\in\{1,4\}$,
and $PCS_2^N$ exist iff $N\in\{1,2,4,8,10,16,20,26,32,34,40,50\}$.
\end{proposition}

\section{The case $p=3$}

If $PCS_3^N$ exist and $N>1$ then (\ref{red}) implies that $N$ is
divisible by 4. Explicit examples of $PCS_3^N$ for $N=4,8,12$
and 16 are given in \cite{BA2}. The non-existence of $PCS_3^{20}$
was first established by a computer search in \cite{BA2} and then
theoretically in \cite{AX}. In our previous note \cite{DZ1}
we have constructed $PCS_3^N$ for $N=24,28$ and 32. We shall
now give the SDS's with parameters
\begin{eqnarray*}
&& (36;15,15,15;18),\quad (40;19,18,15;22), \\
&& (44;20,20,17;24),\quad (48;24,24,18;30),
\end{eqnarray*}
Since these parameter sets satisfy the condition (\ref{red}),
the facts mentioned in section \ref{SDS} imply that $PCS_3^N$
exist for $N=36,40,44$ and 48. 

\newpage

The four SDS's are:
\begin{eqnarray*}
&& N=36: \\
&& X_1=\{0,1,2,3,4,6,7,11,13,15,18,21,23,27,29\}, \\
&& X_2=\{0,1,2,6,7,8,10,11,13,14,18,23,26,27,29\}, \\
&& X_3=\{0,1,3,4,6,7,8,13,14,15,18,21,23,27,32\}; \\
&& N=40: \\
&& X_1=\{0,2,3,4,5,6,7,9,10,12,14,18,19,20,24,28,31,33,34\}, \\
&& X_2=\{0,2,3,4,5,8,9,12,14,15,20,21,22,25,27,29,31,35\}, \\
&& X_3=\{0,1,2,3,7,8,10,11,14,18,19,22,25,27,30\}; \\
&& N=44: \\
&& X_1=\{0,1,2,3,4,5,6,9,11,12,16,17,19,23,24,25,28,32,35,39\}, \\
&& X_2=\{0,2,3,4,5,7,8,12,13,14,17,18,19,21,27,29,31,34,37,40\}, \\
&& X_3=\{0,1,4,5,6,7,9,13,14,16,19,24,25,27,31,33,35\}; \\
&& N=48: \\
&& X_1=\{0,1,2,5,6,7,8,12,13,14,15,18,20,23,25,27,28,29,33,36, \\
&& 37,39,41,44\}, \\
&& X_2=\{0,1,2,3,4,5,6,10,11,13,16,17,19,20,22,25,26,28,29,30, \\
&& 32,34,36,37\}, \\
&& X_3=\{0,1,4,5,7,9,10,11,15,18,19,20,22,27,29,35,38,45\}.
\end{eqnarray*}

Hence, we have
\begin{proposition} \label{slu-3}
In the range $N\le50$, $PCS_3^N$ exist iff
\[ N\in\{1,4,8,12,16,24,28,32,36,40,44,48\}. \]
\end{proposition}

\section{The case $p=5$}

If $PCS_5^N$ exist and $N>1$ then (\ref{red}) implies that $N$ is
divisible by 4. Clearly $PCS_5^N$ exist if $PCS_2^N$ and $PCS_3^N$
exist. Hence, it remains to consider the  cases 
$N=12,20,24,28,36,44$ or 48.
A $PCS_5^{12}$ is given explicitly in \cite{BA2}.
In our previous note \cite{DZ1} we have constructed
a $PCS_5^N$ for $N=20,24,28$ and 36. We shall
now give the SDS's with parameters
\[ (44;21,20,19,18,17;40),\quad (48;23,21,21,20,19;44). \]
Since these parameter sets satisfy the condition (\ref{red}),
the existence of $PCS_5^N$ is established for $N=44$ and 48.
The two SDS's are:
\begin{eqnarray*}
&& N=44: \\
&& X_1=\{0,1,2,3,4,5,7,8,10,11,14,15,19,22,24,26,29,31,32,37,38\}, \\
&& X_2=\{0,2,3,4,5,6,8,10,13,14,18,19,21,24,25,30,31,34,38,40\}, \\
&& X_3=\{0,2,3,4,6,7,10,13,15,19,20,21,27,28,30,32,35,37,39\}, \\
&& X_4=\{0,1,3,4,5,7,9,11,12,15,17,20,21,26,30,31,32,33\}, \\
&& X_5=\{0,1,2,3,5,6,9,10,11,13,14,19,24,25,32,34,37\}; \\
&& N=48: \\
&& X_1=\{0,1,2,3,4,5,8,10,11,13,18,19,20,22,24,25,26,28,30,35, \\
&& \quad 38,40,45\}, \\
&& X_2=\{0,1,2,3,4,7,9,10,13,17,18,19,21,22,26,27,30,32,33,36, \\
&& \quad 41\}, \\
&& X_3=\{0,1,3,5,6,7,8,9,11,12,13,15,19,20,21,30,31,34,35,41,42\}, \\
&& X_4=\{0,1,2,4,5,10,13,14,16,17,19,21,26,27,29,31,34,36,37,40\}, \\
&& X_5=\{0,2,4,6,8,9,12,13,15,19,20,24,25,26,29,32,36,41,43\}.
\end{eqnarray*}

Thus we have
\begin{proposition} \label{slu-5}
In the range $N\le50$, $PCS_5^N$ exist iff $N$ is $1$ or a multiple of $4$.
\end{proposition}

\section{The case $p=6$}

If $PCS_6^N$ exist and $N>1$ then (\ref{red}) implies that $N$ is
even. Clearly $PCS_5^N$ exist if $PCS_2^N$ or $PCS_3^N$ exist.
Hence, it remains to consider the  cases 
6,14,18,22,30,38,42 and 46.
As mentioned in \cite{BA2}, a $PCS_6^6$ can be constructed by using
the rows of a 6 by 6 perfect binary array. Such array has been
constructed in \cite{BA1}.

In our previous note \cite{DZ1} we have constructed
a $PCS_6^N$ for $N=14,18,22$ and 30. We shall
now give the SDS's with parameters
\begin{eqnarray*}
&& (38;18,17,16,16,16,14;40),\quad (42;19,18,18,18,17,17;44),\\
&& (46;21,21,21,21,21,16;52).
\end{eqnarray*}
Since these parameter sets satisfy the condition (\ref{red}),
the existence of $PCS_6^N$ is established for $N=38,42$ and 46.
The four SDS's are:
\begin{eqnarray*}
&& N=38: \\
&& X_1=\{0,1,2,6,7,10,11,12,13,17,18,20,21,22,25,27,29,33\}, \\
&& X_2=\{0,1,3,4,7,8,9,10,13,14,16,19,21,22,26,27,29\}, \\
&& X_3=\{0,1,2,3,4,7,8,9,11,16,18,19,21,22,25,31\}, \\
&& X_4=\{0,1,2,4,6,8,10,12,15,17,18,23,25,26,31,34\}, \\
&& X_5=\{0,2,3,4,5,6,9,14,16,18,20,23,26,27,30,35\}, \\
&& X_6=\{0,1,3,4,5,10,11,13,14,15,18,23,25,29\}; \\
&& N=42: \\
&& X_1=\{0,1,2,3,5,6,7,9,12,14,17,18,19,24,26,28,32,34,39\}, \\
&& X_2=\{0,1,2,3,4,5,7,10,15,17,18,19,21,25,26,27,31,37\}, \\
&& X_3=\{0,1,2,6,9,10,12,13,14,15,18,20,21,23,27,28,30,37\}, \\
&& X_4=\{0,2,4,5,6,8,9,11,15,16,18,19,22,25,27,29,33,35\}, \\
&& X_5=\{0,1,2,4,5,9,13,14,17,18,20,23,24,25,30,33,36\}, \\
&& X_6=\{0,1,2,4,7,8,9,12,13,16,20,21,22,23,26,31,34\}; \\
&& N=46: \\
&& X_1=\{0,1,2,3,5,6,8,9,12,14,15,18,20,24,26,27,32,34,36,37,41\}, \\
&& X_2=\{0,1,2,3,5,7,8,9,12,14,18,19,21,22,25,28,29,30,32,34,35\}, \\
&& X_3=\{0,2,3,4,7,8,9,10,11,14,15,16,19,20,24,25,28,33,35,36, \\
&& \quad 40\}, \\
&& X_4=\{0,1,2,3,5,8,10,13,14,16,18,20,22,23,24,27,29,31,32,38, \\
&& \quad 41\}, \\
&& X_5=\{0,1,2,3,5,6,9,11,13,14,15,17,18,21,24,25,29,36,38,40, \\
&& \quad 41\}, \\
&& X_6=\{0,1,2,3,4,10,11,15,18,20,21,26,28,30,33,34\}. \\
\end{eqnarray*}

Thus we have
\begin{proposition} \label{slu-6}
In the range $N\le50$, $PCS_6^N$ exist iff $N=1$ or $N$ is even.
\end{proposition}

In the case $N=42$ we have found another non-equivalent SDS
with the same parameter set:
\begin{eqnarray*}
&& N=42: \\
&& X_1=\{0,1,3,4,6,8,9,11,12,14,15,17,19,21,23,28,29,31,38\}, \\
&& X_2=\{0,1,2,3,4,5,9,12,13,14,15,19,20,24,28,29,33,35\}, \\
&& X_3=\{0,1,2,4,5,6,7,9,14,15,20,22,25,26,29,32,34,36\}, \\
&& X_4=\{0,1,2,4,5,7,9,10,13,17,19,20,25,26,30,31,33,37\}, \\
&& X_5=\{0,1,2,3,4,7,8,11,14,19,20,24,27,28,30,32,35\}, \\
&& X_6=\{0,1,2,3,4,7,10,12,13,16,17,19,21,23,25,36,37\}.
\end{eqnarray*}

\section{Conclusion}

We reconsider the problem of constructing periodic complementary
sequences $PCS_p^N$ ($p$ sequences, each of length $N$) in
the range $p\le12$, $N\le50$. This problem has been addressed
by B\"omer and Antweiler in their paper \cite{BA2}, where
they presented a diagram showing for which pairs $(p,N)$
they were able to construct such set of sequences. Many
cases were left as undecided. The non-existence
was established in a number of cases. Subsequently we have
reduced the number of undecided cases to just 26, see \cite{DZ2}.
For various methods of constructing $PCS_p^N$ one should also
consult the paper \cite{FSX}.

In the present note we have eliminated all 26 undecided cases by 
constructing suitable supplementary difference sets.
Table 1 shows how one can construct a $PCS_p^N$ in
the range $p\le12$ and $N\le50$ when one exists.
By Proposition \ref{slu-4}, if $p$ is divisible by 4 then
$PCS_p^N$ exist for all $1\le N\le50$.
Therefore we omit the rows with $p$ divisible by 4.
When $p$ is not divisible by 4 and $N>1$,
then the equation (\ref{red}) implies that $PCS_p^N$ may exist
only for $N$ even. For this reason we omit the odd $N$'s from
the table.

A blank entry in position $(p,N)$ means that $PCS_p^N$ do not
exist. A bullet entry means that a $PCS_p^N$ exists and has to
be constructed directly by using a well known technique
such as one for Golay pairs, perfect binary arrays,
or an SDS, etc. The references for the bullet entries $(p,N)$
are given in the main text.
The circle entry means that a $PCS_p^N$ can be constructed in
a trivial way, i.e., by combining several $PCS_q^N$ for $q<p$
corresponding to bullet entries or $q=4$.

\begin{landscape}
\[ \begin{array}{rcccccccccccccccccccccccccc}
\multicolumn{27}{c}{\textbf{Table 1: Construction of $PCS_p^N$ for
even $N\le50$}} \\
\\ \hline \\
&&2&4&6&8&10&&&&&20&&&&&30&&&&&40&&&&&50\\
\\ \hline \\
1&&&\bt&&&&&&&&&&&&&&&&&&&&&&&\\
2&&\bt&\kr&&\bt&\bt&&&\bt&&\bt&&&\bt&&&\bt&\bt&&&\bt&&&&&\bt\\
3&&&\kr&&\bt&&\bt&&\bt&&&&\bt&&\bt&&\bt&&\bt&&\bt&&\bt&&\bt&\\
5&&&\kr&&\kr&&\bt&&\kr&&\bt&&\bt&&\bt&&\kr&&\bt&&\kr&&\bt&&\bt&\\
6&&\kr&\kr&\bt&\kr&\kr&\kr&\bt&\kr&\bt&\kr&\bt&\kr&\kr&\kr&\bt&\kr
&\kr&\kr&\bt&\kr&\bt&\kr&\bt&\kr&\kr\\
7&&&\kr&&\kr&&\kr&&\kr&&\kr&&\kr&&\kr&&\kr&&\kr&&\kr&&\kr&&\kr&\\
9&&&\kr&&\kr&&\kr&&\kr&&\kr&&\kr&&\kr&&\kr&&\kr&&\kr&&\kr&&\kr&\\
10&&\kr&\kr&\kr&\kr&\kr&\kr&\kr&\kr&\kr&\kr&\kr&\kr&\kr&\kr&\kr
&\kr&\kr&\kr&\kr&\kr&\kr&\kr&\kr&\kr&\kr\\
11&&&\kr&&\kr&&\kr&&\kr&&\kr&&\kr&&\kr&&\kr&&\kr&&\kr&&\kr&&\kr&\\
\end{array} \]
\end{landscape}


\begin{thebibliography}{99}

\bibitem{TA}
T.H. Andres, Some combinatorial properties of complementary
sequences, M.Sc. Thesis, University of Manitoba, Winnipeg, 1977.

\bibitem{AX}
K.T. Arasu and Q. Xiang, On the existence of periodic complementary
binary sequences,
Designs, Codes and Cryptography {\bf 2} (1992), 257--262.

\bibitem{DA}
D. Ashlock, Finding designs with genetic algorithms, in W.D. Wallis
(Ed.), Computational and Constructive Design Theory, pp. 49--65,
Kluwer Academic Publishers, Dordrecht/Boston/London, 1996.

\bibitem{BA1}
L. B\"omer and M. Antweiler, Two-dimensional perfect binary arrays
with 64 elements,
IEEE Trans. Inform. Theory {\bf 36} (1990), 411--414.

\bibitem{BA2}
\bysame, Periodic complementary binary sequences,
IEEE Trans. Inform. Theory {\bf 36} (1990), 1487--1494.

\bibitem{BF}
P.B. Borwein and R.A. Ferguson, A complete description of Golay
pairs for lengths up to 100,
Math. Comp. {\bf 73} (2003), 967--985.

\bibitem{HCD}
C.J. Colbourn and J.H. Dinitz, Editors, Handbook of Combinatorial Designs,
2nd edition, Chapman \& Hall, Boca Raton/London/New York, 2007.

\bibitem{DZ1}
D.\v{Z}. \DJo{},
Survey of cyclic $(v;r,s;\la)$ difference families with $v\le50$,
Facta Universitatis (Ni\v{s}), Ser. Mathematics and Informatics
{\bf 12} (1997), 1--13.

\bibitem{DZ2}
\bysame. Note on periodic complementary sets of binary sequences,
Designs, Codes and Cryptography {\bf 13} (1998), 251--256.

\bibitem{DZ3}
\bysame, Equivalence classes and representatives of Golay sequences,
Discrete Math. {\bf 189} (1998), 79--93.

\bibitem{DZ4}
\bysame, Aperiodic complementary quadruples of binary sequences,
J. Combin. Math. Combin. Comp. {\bf 27} (1998), 3--31.
Correction, ibid {\bf 30} (1999), 254.

\bibitem{DZ5}
\bysame, Cyclic $(v;r,s;\la)$ difference families with two base
blocks and $v\le50$,
preprint (2007), 39 pp.

\bibitem{FSX}
K. Feng, P.J-S. Shiue, and Q. Xiang, On aperiodic and periodic
complementary binary sequences,
IEEE Trans. Inform. Theory {\bf 45} (1999), 296--303.

\bibitem{BS}
B. Schmidt, Cyclotomic integers and finite geometry,
J. Amer. Math. Soc. {\bf 12} (1999), 929--952.

\bibitem{RT}
R.J. Turyn, Hadamard matrices, Baumert--Hall units, four-symbol
sequences, pulse compression and surface wave encodings,
J. Combin. Theory A {\bf 16} (1974), 313--333.

\bibitem{CY}
C.H. Young, Maximal binary matrices and sum of two squares,
Math. Comp. {\bf 30} (1976), 361--366.

\end{thebibliography}
\end{document}